\documentclass[showpacs,twocolumn,prl]{revtex4}
\usepackage{amsmath}
\usepackage{amssymb}
\usepackage{graphicx}
\usepackage{dcolumn}
\usepackage{bm}

\input{tcilatex}
\begin{document}

\title{Electric field modulation of topological order in thin film
semiconductors }
\author{Zhan-Feng Jiang$^{\ast }$, Rui-Lin Chu$^{\ast }$, and Shun-Qing Shen$%
^{\dag }$}
\affiliation{Department of Physics, and Center of Theoretical and Computational Physics,
The University of Hong Kong, Pokfulam Road, Hong Kong}
\date{\today}

\begin{abstract}
We propose a method that can consecutively modulate the topological orders
or the number of helical edge states in ultrathin film semiconductors
without a magnetic field. By applying a staggered periodic potential, the
system undergoes a transition from a topological trivial insulating state
into a non-trivial one with helical edge states emerging in the band gap.
Further study demonstrates that the number of helical edge state can be
modulated by the amplitude and the geometry of the electric potential in a
step-wise fashion, which is analogous to tuning the integer quantum Hall
conductance by a megntic field. We address the feasibility of experimental
measurement of this topological transition.
\end{abstract}

\maketitle

Edge and surface physics arising from the topological insulators have been a
major research focus in the condensed matter physics recently.\cite%
{Haldane-88prl,Kane-05prl,Bernevig-06science,Konig-07science,Fu-07prl,Hsieh-08nature,Xia-09np,Zhang-09np,Chen-09science}
Usually spin-orbit coupling in these materials is so strong that the band
gap between the conduction and valence bands can be inverted. In this case
edge states or surface states emerge in the band gap, protected by
time-reversal symmetry. The $Z_{2}$ invariant is established to govern the
topological properties of both two- and three-dimensional (3D) topological
insulators.\cite{Kane-05prl-b,Moore-07prb} Searching for this topological
state of quantum matter has been focused on specific materials such as
HgTe/CdTe quantum wells\cite{Konig-07science}, Bismuth alloys\cite%
{Hsieh-08nature,Xia-09np,Zhang-09np,Chen-09science} and transition metal
oxide Na$_{2}$IrO$_{3}$.\cite{Shitade-09prl} It was observed that there are
five surface states in Bi$_{1-x}$Sb$_{x}$\cite{Hsieh-08nature} and single
Dirac cone in Bi$_{2}$Se$_{3}$\cite{Xia-09np} and Bi$_{2}$Te$_{3}$\cite%
{Chen-09science}while there is only a pair of helical edge states in
HgTe/CdTe quantum wells.\cite{Konig-07science,Roth-09science} Up to now
there is neither experimental report nor theoretical proposal to tune the
topological number in these system in a controllable way. It would be both
theoretically and experimentally interesting if the topological number or
the number of the helical edge states or surface states can be modulated
like the quantum Hall conductance in a magnetic field.

In this paper, we propose a feasible approach to modulate consecutively the
pair number of helical edge states by an electric means. To be specific, we
consider a quasi two dimensional semiconductor thin film or quantum well
with strong spin-orbit coupling, which can be either topologically trivial
or non-trivial. We apply a lateral surface superlattice (SSL) that creates a
periodic potential on the film. It is found that the system changes from a
trivial insulator into a quantum spin Hall (QSH) insulator with helical edge
states emerging at the sample edges. This phenomenon can be understood
intuitively as the periodic potential splits the band structure into
multi-mini-bands, inverts the band gap and changes the $Z_{2}$ order of the
bulk. However, unlike the QSH insulator, the band inversion story doesn't
stop here. By further increasing the gate geometry or potential magnitude,
the pair number of helical edge states goes up step wise while the bulk band
gap remains finite. In this manner we create a topological insulator whose
edge states are tunable by purely electrical means.

We start with an effective four-band Hamiltonian with the time-reversal
invariance,%
\begin{equation}
H_{0}=\left( 
\begin{array}{cc}
h_{+}(-i\partial _{x},-i\partial _{y}) & 0 \\ 
0 & h_{-}(-i\partial _{x},-i\partial _{y})%
\end{array}%
\right)
\end{equation}%
where $h_{\pm }=+D(\partial _{x}^{2}+\partial _{y}^{2})+A(-i\partial
_{x}\sigma _{y}+i\partial _{y}\sigma _{x})\pm \lbrack \Delta /2+B(\partial
_{x}^{2}+\partial _{y}^{2})]\sigma _{z}$ and $\sigma _{\alpha }$ are the
Pauli matrices. The model was first introduced for the HgTe/CdTe quantum
well for QSH effect by Bernevig, Hughes and Zhang,\cite{Bernevig-06science}
and recently was derived for an ultra-thin film of 3D topological insulator.%
\cite{Lu-09xxx} The two cases have different basis although the forms are
identical. $h_{\pm }$ consists of the invariants in the irreducible
representation $D_{1/2}$ of SU(2) group.\cite{Winkler} The whole model $%
H_{0} $ keeps the time reversal invariance. %  The
% model can decribe the topological quantum phase transition between a
% topologically trivial band insulator to a non-trial one.\cite%
% {Bernevig-06science} 
$h_{+}$ and $h_{-}$ are the "time" reversal counterparts under the operation 
$\Theta =-i\sigma _{y}K$ where $K$ is the complex conjugate operator, $%
h_{+}=\Theta ^{-1}h_{-}\Theta $. The block diagonalized form in $H_{0}$
allows us to study $h_{+}$ and $h_{-}$ separately and then put together to
gain the physics for $H_{0}$. Additional terms such as those of bulk or
structure inversion asymmetry can couple them together. Here we first assume
that these effects are weak and negligible.

The idea of a lateral SSL dates back to the 1970s. Various alternative
techniques are known capable of creating SSLs.\cite{Tsu-05} It is known that
periodic potential created by SSLs induces Bloch mini-bands and mini-gaps on
a 2DEG, which results in interesting transport behaviors. Consider this
model on a SSL. To be specific and without loss of generality, we introduce
a square-wave-shaped periodic potential, 
\begin{equation}
V(y)=\left\{ 
\begin{array}{c}
V_{0},\text{ }0\leqslant y<d/2 \\ 
-V_{0},-d/2\leqslant y<0%
\end{array}%
\right. 
\end{equation}%
and $V(y+d)=V(y)$. We write the Fourier series for $V(y)$ as $%
V(y)=\sum\limits_{n}V_{n}e^{inQy}$ with the reciprocal vector $Q=2\pi /d$
and the Fourier transform component $V_{n}=2iV_{0}/n\pi $ ($n=\pm 1,\pm
3,\cdots $.). According to the Bloch theorem,\cite{Kittel} the single
particle wave-function of $h_{+}$ for the band $\epsilon _{n,k}^{s}$ in this
periodic potential $V(y)$ has the form $\Psi
_{n}^{s}(x,y)=e^{i(k_{x}x+k_{y}y)}u_{n,k_{y}}^{s}(y)$ with $%
u_{n,k_{y}}^{s}(y)=u_{n,k_{y}}^{s}(y+d)$ where $k_{y}$ is confined in the
first Brillouin zone (BZ) $k_{y}\in \lbrack -Q/2,+Q/2]$, $n$ is an integer,
and the superscript $s(=c,v)$ denotes the conduction ($c$) and valence ($v$)
band. The wave function can be expressed as $u_{n,k_{y}}^{s}(y)=%
\sum_{m}C_{n,k_{y}}^{s}(m)e^{imQy}$, with the coefficients $C_{n,k_{y}}$
determined by the central equation%
\begin{equation}
\lbrack h_{+}(k_{x},mQ+k_{y})-\epsilon
_{n,k}^{s}]C_{n,k_{y}}^{s}(m)+\sum_{l}V_{l}C_{n,k_{y}}^{s}(m-l)=0
\label{central}
\end{equation}%
with the condition $C_{n,k_{y}}^{s}(m)=\delta _{n,m}C_{n,k_{y}}^{s}(n)$ for $%
V=0$. In this way, the electron band is folded into many mini-bands confined
in the first BZ as illustrated schematically in Figs.1(a) and (b), where the
labels ($n,s$) represent different mini-bands.

\begin{figure}[t]
\begin{center}
\includegraphics[scale=0.65]{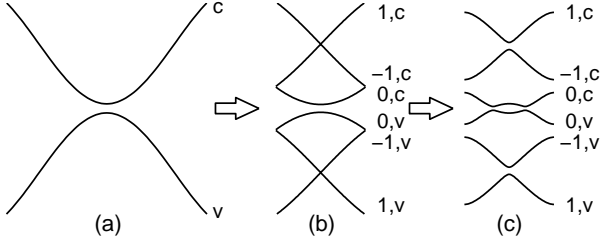}
\end{center}
\caption{Evolution of energy bands in a modulated periodic potential. (a)
The band in the absence of periodic potential; (b) Mini-bands formed in a
periodic potential; (c) Shifting of the minibands leads to the band
invertion and band anti-crossing.}
\label{fig1}
\end{figure}

In the presence of $V$, the mini-band structure can be obtained numerically
by solving the central equation. It is found that when the potential $V$
increases the conduction mini-bands shift down and the valence mini-bands
shift up, eventually the two bands cross and form a inverted band gap as
shown in Fig. 1(c). Anticrossing between the two bands occurs, which is
induced by the interaction term $A$. A simple perturbative viewpoint helps
us to apprehend why the periodic potential $V$ tends to induce the band
inversion. Take the band ($0,c$) for our illustration. For a weak $V$, the
calculation up to the second order perturbation gives the energy correction
for the band ($0,c$) 
\begin{equation}
\Delta \epsilon _{0,k}^{c}=\underset{m,s,k_{y}^{\prime }}{{\displaystyle\sum 
}}\frac{\left\vert \left\langle u_{0,k_{y}}^{c}\left\vert V(\mathbf{r}%
)\right\vert u_{m,k_{y}^{\prime }}^{s}\right\rangle \right\vert ^{2}}{%
\epsilon _{0,k}^{c}-\epsilon _{m,k_{y}^{\prime }}^{s}},
\end{equation}%
here the summation excludes the band (0, c) itself, $u_{m,k_{y}^{\prime
}}^{s}$ and $\epsilon _{n,k}^{s}$ are the unperturbated eigen state and
eigen energy of the mini-band in the first BZ. Using the Fourier series of
the potential, we have $\left\langle u_{0,k_{y}}^{c}\left\vert V(\mathbf{r}%
)\right\vert u_{m,k_{y}^{\prime }}^{s}\right\rangle =V_{-m}\delta
_{k_{y},k_{y}^{^{\prime }}}\left\langle
u_{0,k_{y}}^{c}|u_{m,k_{y}}^{s}\right\rangle .$ $V(\bm{r})$ couples the band
($0,c$) and other conduction bands ($n,s=c$) stronger than the valence bands
since they come from same original bands and share the same spin indices. We
have $\left\vert \left\langle u_{0,k_{y}}^{c}|u_{m,k_{y}}^{c}\right\rangle
\right\vert \gg \left\vert \left\langle u_{0,k_{y}^{\prime
}}^{c}|u_{m,k_{y}}^{v}\right\rangle \right\vert $. Thus we draw the
conclusion that the perturbative correction is always negative, $\Delta
\epsilon _{0,k}^{c}<0$, whose effect is shifting the band ($0,c$) downward.
Similarly, the valence band ($0,v$) shifts upward in the field. At a certain
value of $V$, the two bands cross and open a gap again, \textit{i.e.} a
negative one. Similar argument can be applied to other mini-bands. With
increasing $V$, more and more mini-bands will cross near the zero energy
point. Thus the band inversion occurs consecutively because of the shifting
of the mini-bands from the conduction and valence bands.

The formation of the inverted band gap is of topological and experimental
interests. To gain a quantitative insight of this physical picture, we
discretize the Hamiltonian $h_{+}(\bm{k})$ in Eq. (2) on a square lattice.
Without losing generality, we make the parameters material independent by
setting $A=-B=1$, $D=0$ and the lattice space $a=1$. Eigen energies and
eigen states are solved numerically for the periodic system. Following
Thouless and his co-workers,\cite{Thouless-82prl} we come to calculate the
Hall conductance of $h_{+}$ in the band gap, which is equivalent to the
Chern number of the filled valence bands. The Berry curvature for each band
is defined as\cite{Chang-08jpcm} 
\begin{equation}
\Omega _{n,s}(k)=i\left( \left\langle \frac{\partial u_{n,k}^{s}}{\partial
k_{x}}\right\vert \left. \frac{\partial u_{n,k}^{s}}{\partial k_{y}}%
\right\rangle -\left\langle \frac{\partial u_{n,k}^{s}}{\partial k_{y}}%
\right\vert \left. \frac{\partial u_{n,k}^{s}}{\partial k_{x}}\right\rangle
\right) ,
\end{equation}%
The first Chern number in the band gap is then computed by summing up Berry
curvatures of the occupied bands and integrate over the first BZ $C=\frac{1}{%
2\pi }\sum_{n}\iint d^{2}k\cdot \Omega _{n,v}(k).$ The Hall conductance for $%
h_{+}$ is related to the first Chern number by $\sigma _{xy}=\frac{e^{2}}{h}C
$.\cite{Thouless-82prl,Kohmoto-85ap} In the absence of the potential, the
Chern number was found to be $C=-[sgn(\Delta )+sgn(B)]/2$.\cite{Lu-09xxx}
The signs of the two model parameters in the term $(\Delta /2-Bk^{2})\sigma
_{z}$ of $h_{+}$ determine the value of $C$, $0$ or $\pm 1$, and further
whether the system is topologically trivial or not. In the presence of $V$,
it is found that the Chern number changes by 1 when the valence and
conduction bands cross. By continually changing the potential period $d$ and
potential magnitude $V_{0}$, we obtain a phase diagram for the Chern number
in Fig. 2 with the numbers in the boxes indicating the values. We see that
in this system the Chern number is a function of the potential period $d$
and amplitude $V_{0}$.

\begin{figure}[t]
\begin{center}
\includegraphics[scale=0.6]{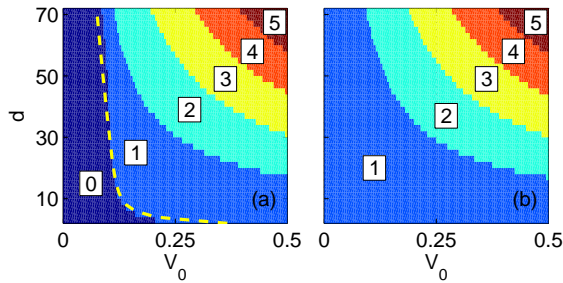}
\end{center}
\caption{(Color online) The phase diagram of Chern number in ($V_{0},d$)
plane for the model with the Fermi energy $E_{f}=0$, (a) $\Delta /2=0.01$;
(b) $\Delta /2=-0.01$;. The dashed line in (a) is obtained by solving Eq.(2)
in the truncation approximation for $m=0$ and $\pm 1$ in $C_{m,k_{y}}^{s}$.
Other parameters are $A=-B=1$, $D=0$ and the lattice space $a=1$.}
\label{fig2}
\end{figure}

Examination of the evolution of the band structure and Berry curvature at
the transition courses further confirms our claim. In Fig. 3 we show the
first two transitions from $C=0$ to $1$, and from $C=1$ to $2$. By
increasing $V_{0}$, the band gap closes and re-opens in Fig. 3(a) and 3(b).
Correspondingly, the Berry curvature peak become sharp, and reverses its
value dramatically, which accounts for the change in the Chern Number. At
the corresponding $\bm{k}$ points, the eigen wave functions also switch
their band indexes quickly. By further increasing $V_{0}$, the peak of the
Berry curvature splits into two sub-peaks, and the third peak with an
opposite sign grows up while the Chern number remain to be $C=1$. In the
second transition, the third peaks reverses again just like the first
transition, and the Chern number changes from $C=1$ to $2$. Therefore
accompanied by every "closing and re-opening" of the band gap the Chern
number changes by one.

\begin{figure}[t]
\begin{center}
\includegraphics[scale=0.6]{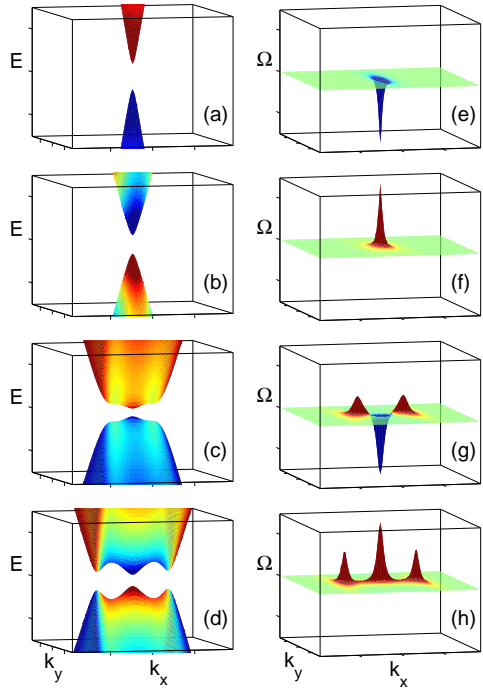}
\end{center}
\caption{(Color online) (a)-(d) Energy dispersion for the first valence and
conduction band for various periodic potential strengths (a) $V_{0}=0.06$;
(b) $0.14$; (c) $0.23$; (d) $0.26$. Color red indicates the dominant eigen
state is spin-up, color blue indicates the dominant eigen state is
spin-down. The other parameters are $\Delta /2=0.01,d=30.$ (a) and (b)
corresponds to the transition of $C=0\rightarrow 1$. (c) and (d) corresponds
to the transition of $C=1\rightarrow 2$. (e)-(h) Berry curvatures of the
valence band around $\Gamma $ point corresponding to the band structures
(a)-(d), respectively.}
\label{fig3}
\end{figure}

\begin{figure}[t]
\begin{center}
\includegraphics[scale=0.65]{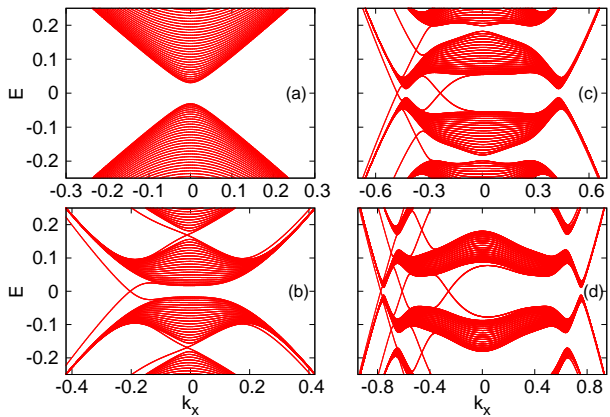}
\end{center}
\caption{Band stucture evolution of a strip geometry in a periodic
potential. Strip width $400$, $d=20$, $\Delta /2=0.03$. (a) $V=0,C=0$; (b) $%
V_{0}=0.3,C=1$; (c) $V_{0}=0.6,C=2$; (d) $V_{0}=1.0,C=3$. Asymmetric
dispersion for edge state originates from the symmetric breaking of the
potential on the stripe with open boundary condition.}
\label{fig4}
\end{figure}

According to the edge-bulk correspondence\cite{Hatsugai-93prl,Qi-06prb}, the
first Chern number determines number of the edge states when the system has
an open boundary. We take a strip geometry that is parallel to the x
direction with open boundary condition along the y direction. We start from
a topological trivial case of $\Delta >0$ and $B<0,$ and scrutinize its
changing in the band structure when varying the periodic potential $V_{0}$.
In the absence of the potential $V$ the system is insulating and
topologically trivial indicated by a positive band gap. As $V_{0}$
increases, the conduction bands shift down and valence bands shift up.
Eventually the two bands cross, and a band gap re-opens. A pair of edge
states appears connecting the valence and conduction bands as shown in Fig.
4b. Detailed analysis shows that the wave functions of these states indeed
reside on the sample edges only. As $V_{0}$ goes further up, more bands
cross and more pairs of edge states are formed at distinct $\bm{k}$ points
as shown in Figs. 4c and 4d. It is interesting to observe that the band gap
at $E=0$ always retains a finite value at certain stages throughout the
evolution. We can see the appearance of the edge states corresponds to the
change of the Chern number from $C=0$ to $C=3$ by comparing Figs. 3 and 4.
Thus modulation of edge state numbers is clearly reflected in the step wise
changing of the Chern numbers.

$h_{+}$ and $h_{-}$ are the time reversal counterparts of each other. Each
edge state $\left\vert \psi _{+}\right\rangle $ in $h_{+}$ has a couterpart $%
\left\vert \psi _{-}\right\rangle =\Theta \left\vert \psi _{+}\right\rangle $
in $h_{-}$, and they form a pair of helical edge state. As a result, the
large Chern number ($C>1$)\ indicates multiple pairs of helical edge states
in the system $H_{0}$ in this periodic potential $V$. For a thin film
fabricated on the substrate, the top-bottom symmetry will be broken due to
the interface of the thin film and the substrate. This fact will remove the
degeneracy of the spectra from $h_{+}$ and $h_{-}$. An off-diagonal term of
structural inversion asymmetry is added to the model $H_{0}$ in Eq. (1),\cite%
{Xue-09np} $\Delta V=V_{SIA}\left( 
\begin{array}{cc}
0 & \sigma _{0} \\ 
\sigma _{0} & 0%
\end{array}%
\right) $ which connects the up and down blocks ($\sigma _{0}$ is the $%
2\times 2$ identity matrix). Similar physics happens if the sample breaks
the bulk inversion symmetry in quantum well.\cite{Konig-08JPSJ} In the
presence of $V_{SIA}$, numerical calculation still demonstrates existence of
the edge states in the four-band model, which is characteristic of QSH
phase. Thus this term does not destroy the QSH phase explicitly.

Feasibility of experimental realization of this phenomenon depends on spin
decoherence length in the sample, and the fabrication of the periodic
potential. The spin decoherence length was estimated to be $1\sim 2\mu m$ in
the HgTe/CdTe quantum well,\cite{Konig-07science,Roth-09science} and
expected to be longer in a thin film of Bi$_{2}$Se$_{3}$.\cite{Xue-09np} On
the periodicity of the modulated electric field, the sub-100nm period of SSL
fabrications was reported a decade ago,\cite{Messica-97prl} and the 3nm
periodicity for a graphene on Ru(111) recently.\cite{Marchini-07prb} For a
set of realistic parameters for HgTe/CdTe quantum well with a normal band
gap $\Delta =4$meV,\cite{Bernevig-06science} the magnitude of the potential
with periodicity $d=50$nm is calculated to $V_{0}=40$meV for the transition
of $C=0$ to $1$ and $V_{0}=110$meV for the transition of $C=1$ to $2$. We
speculate that the modern techniques of superlattice make this electric
field modulation possible.

In summary, the band structure of a thin film or quantum well are folded
into the mini-bands in the reduced BZ by a periodic potential, and can be
modulated such that the conduction bands shift downward and the valence
bands shift upward. Each process of the band gap closing and re-opening will
change the Chern number by one. As a result the number of the helical edge
states will increase or decrease by one. This demonstrates the possibility
of the electric field modulation of topological orders in the thin film
semiconductors, which is analogous to the integer quantum Hall effect in a
strong magnetic field. Direction of electrons of the helical edge states is
of interests in quantum information and quantum processing. Controllable
number of the helical edge states will pave an alternative route for
application of edge state physics in the future.

This work was supported by the Research Grant Council of Hong Kong under
grant No.: HKU 704107 and HKU 10/CRF/08.

$^{\ast }$These authors contributed equally to this work.

$^{\dag }$E-mail: sshen@hkucc.hku.hk


\begin{thebibliography}{99}
\bibitem{Haldane-88prl} For introductions to topological insulators, see S.
C. Zhang, Physics 1, 6 (2008); M. Buttiker, Science 325, 278 (2009); J.
Moore, Nature Physics 5, 378 (2009).

\bibitem{Kane-05prl} C. L. Kane and E. J. Mele, Phys. Rev. Lett. 95, 226801
(2005)

\bibitem{Bernevig-06science} B. A. Bernevig, T. L. Hughes, and S. C. Zhang,
Science \textbf{314}, 1757 (2006).

\bibitem{Konig-07science} M. K\"{o}nig, S. Wiedmann, C. Br\"{u}ne, A. Roth,
H. Buhmann, L. W. Molenkamp, X. L. Qi, and S. C. Zhang, Science \textbf{318}%
, 766 (2007).

\bibitem{Fu-07prl} Liang Fu, C. L. Kane, and E. J. Mele, Phys. Rev. Lett.
98, 106803 (2007).

\bibitem{Hsieh-08nature} D. Hsieh, D. Qian, L. Wray, Y. Xia, Y. S. Hor, R.
J. Cava and M. Z. Hasan, Nature 452, 970 (2008).

\bibitem{Xia-09np} Y. Xia, D. Qian, D. Hsieh, L. Wray, A. Pal, H. Lin, A.
Bansil, D. Grauer, Y. S. Hor, R. J. Cava and M. Z. Hasan, Nature Physics 5,
398 (2009).

\bibitem{Zhang-09np} H. Zhang, C. X. Liu, X. L. Qi, X. Dai, Z. Fang, and S.
C. Zhang, Nature phys. \textbf{5}, 438 (2009).

\bibitem{Chen-09science} Y. L. Chen, J. G. Analytis, J.-H. Chu, Z. K. Liu,
S.-K. Mo, X. L. Qi, H. J. Zhang, D. H. Lu, X. Dai, Z. Fang, S. C. Zhang, I.
R. Fisher, Z. Hussain, and Z.-X. Shen, Science 325, 178 (2009).

\bibitem{Kane-05prl-b} C. L. Kane and E. J. Mele, Phys. Rev. Lett. 95,
146802 (2005)

\bibitem{Moore-07prb} J. E. Moore and L. Balents, Phys. Rev. B 75 121306
(2007)

\bibitem{Shitade-09prl} A. Shitade, H. Katsura, J. Kune\v{s}, X. L. Qi, S.
C. Zhang, and N. Nagaosa, Phys. Rev. Lett. 102, 256403 (2009).

\bibitem{Roth-09science} A. Roth, C. Brune, H. Buhmann, L. W. Molenkamp, J.
Maciejko, X. L. Qi, S. C. Zhang, Science 325, 294 (2009)

\bibitem{Tsu-05} R. Tsu, \textit{Superlattice to Nanoelectronics} (Elsevier,
Oxford, 2005).

\bibitem{Lu-09xxx} H. Z. Lu, W. Y. Shan, W. Yao, Q. Niu, and S. Q. Shen,
arXiv: 0908.3120.

\bibitem{Winkler} R. Winkler, \textit{Spin-orbit Coupling Effects in
Two-Dimensional Electron and Hole Systems}, p. 64, (Springer, Berlin, 2003).

\bibitem{Kittel} C. Kittel, Introduction to Solid State Physics (7th ed.),
Chap. 7, p. 183, (John Wiley \& Sons, New York, 1996)

\bibitem{Thouless-82prl} D. J. Thouless, M. Kohmoto, M. P. Nightingale, and
M, deb Nijs, Phys. Rev. Lett. 49, 405 (1982).

\bibitem{Chang-08jpcm} M. C. Chang and Q. Niu, J. Phys.: Cond. Mat. 20,
193202 (2008); D. Xiao, M. C. Chang and Q. Niu, arXiv: 0907.2021

\bibitem{Kohmoto-85ap} M. Kohmoto, Ann. Phys. (NY) 160, 343 (1985).

\bibitem{Hatsugai-93prl} Y. Hatsugai, Phys. Rev. Lett. 71, 3697 (1993).

\bibitem{Qi-06prb} X. L. Qi, Y. S. Wu, and S. C. Zhang, Phys. Rev. B 74
045125 (2006)

\bibitem{Xue-09np} Y. Zhang et al, preprint submitted to Nature Phsyics,
(2009); W. Y. Shan et al, in preparation.

\bibitem{Konig-08JPSJ} M. K\"{o}nig, H. Buhmann, L. W. Molenkamp, T. Hughes,
C. X. Liu, X. L. Qi, and S. C. Zhang, J. Phys. Soc. Jpn. \textbf{77}, 031007
(2008).

\bibitem{Messica-97prl} A. Messica, A. Soibel, U. Meirav, A. Stern, H.
Shtrikman, V. Umansky, and D. Mahalu, Phys. Rev. Lett. \textbf{78}, 705
(1997).

\bibitem{Marchini-07prb} S. Marchini, S. Gunther, and J. Wintterlin, Phys.
Rev. B 76, 075429 (2007)
\end{thebibliography}
\end{document}